\documentstyle[12pt,epsf]{article}
\begin{document}
\voffset -3.0cm
\textwidth=16cm
\textheight=22cm
\begin{titlepage}
\pagestyle{empty}
\begin{flushright}
\vbox{
{\bf TTP 93-40}\\
{\rm February 1994}}
\end{flushright}
\renewcommand{\thefootnote}{\fnsymbol{footnote}}
\vspace{0.5cm}
\begin{center}
{\large DISTRIBUTIONS OF LEPTONS
IN DECAYS\\
OF POLARISED HEAVY QUARKS
\footnote{ Work upported in part by
BMFT under contract 056KA93P and by KBN}  }
\end{center}
\vskip1.0cm
\hyphenation{Germany}
\begin{center}
{\sc A. Czarnecki}\\
\vskip0.2cm
Institut f. Physik, Johannes Gutenberg-Univ.,
D-55099 Mainz, Germany\\
\vskip0.7cm
{\sc M. Je\.zabek}
\\
\vskip0.2cm
Institute of Nuclear Physics, Kawiory 26a, PL-30055 Cracow,
Poland \footnote{Permanent address}
\\
and\\
Institut f\"ur Theoretische Teilchenphysik, Universit\"at Karlsruhe\\
D-76128 Karlsruhe, Germany
\end{center}
\vskip1.5cm
\begin{center}
Abstract
\end{center}
{\small
Analytic formulae are given for QCD corrections to the lepton
spectra in decays of polarised up and down type heavy quarks.
These formulae are much simpler than the published ones
for the corrections to the spectra of charged leptons
originating from the decays of unpolarised quarks and polarised
up type quarks.
Distributions of leptons in semileptonic $\Lambda_c$
and $\Lambda_b$ decays can be used as spin analysers for
the corresponding heavy quarks. Thus our results can be applied
to the decays of polarised charm and bottom quarks. For the
charged leptons the corrections to the asymmetries are found
to be small in charm decays whereas for bottom decays they exhibit
a non-trivial dependence on the energy of the charged lepton.
Short life-time  enables polarisation studies for the top quark.
Our results are directly applicable for processes involving
polarised top quarks.
\vskip1.0cm    }

\end{titlepage}
\setcounter{footnote}{0}
\renewcommand{\thefootnote}{\arabic{footnote}}

\section{Introduction}

Inclusive semileptonic decays of heavy flavours play important
role in present day particle physics. In near future with
increasing statistics at LEP and good prospects for $B$-factories
quantitative description of these processes may offer the most
interesting tests of the standard quantum theory
of particles. At the high energy frontier semileptonic decays
of the top quark will be instrumental in establishing its
properties \cite{Kuehn1,Kuehn2}.

Recent theoretical developments,
discovery of Heavy Quark Symmetries \cite{IW,reviews}
and Effective Theory \cite{HQET} as well as subsequent applications
of an expansion in inverse powers of the heavy quark mass $m_Q$
\cite{Bigi}, have led to a consistent  treatment
of semileptonic decays of charmed and beautiful hadrons.
In the limit $m_Q\to\infty$ the inclusive decays of heavy flavour
hadrons are described by the decays of the corresponding
heavy quarks. For finite quark masses there arise non-perturbative
corrections due to the heavy quark motion inside the heavy hadrons
\cite{Alt}. A systematic approach to these corrections
based on $1/m_Q$ expansion is described in \cite{Bigi}
and other articles cited in \cite{reviews}.
In the present article we  consider only perturbative QCD corrections
to the decays of heavy quarks. Non-perturbative effects have to be
treated according to the recipies provided by
the simple model of \cite{Alt} or by the more formal approach
of \cite{Bigi}. On the other hand inclusion of perturbative
corrections is necessary for a quantitative description of
semileptonic decays because these corrections are of the order
of 20 percent.

Interesting new opportunities are provided by decays of polarised
heavy quarks. It is well known \cite{Kuehn1,Kuehn2} that
the heavy quarks produced in $Z^0$ decays are polarised. According
to the Standard Model the degree of longitudinal polarisation
is fairly large, amounting to $\langle P_b\rangle = -0.94$ for $b$
and $\langle P_c\rangle = -0.68$ for $c$ quarks~\cite{Kuehn1}.
The polarisations depend weakly on the production angle.
QCD corrections to the production cross section
have been calculated
recently \cite{KPT}. Unfortunately, this original high polarisation
is to large extend washed out during hadronisation into mesons.
At the time being only charmed and beautiful $\Lambda$ baryons
seem to offer a practical method to measure the polarisation of the
corresponding heavy quarks. This method proposed by Bjorken
\cite{Bjo} has been successfully applied to both semileptonic
\cite{BKKZ}
and hadronic \cite{JRR} decays of Lambda baryons.
The polarisation transfer from a heavy quark $Q$ to the
corresponding $\Lambda_Q$ baryon is 100\% \cite{CKPS},
at least in the limit $m_Q\to\infty$. In view of
the growing sample of $\Lambda_c$ and $\Lambda_b$ baryons
produced at LEP there arises an opportunity to measure the
polarisation of the $c$ and $b$ quarks originating from the
decays of the $Z$ bosons.  The angular distributions of charged
leptons \cite{zerwas,mele92} and neutrinos \cite{cjkk}
from semileptonic decays of $\Lambda_c$ and $\Lambda_b$
can be used as spin analysers for the decaying heavy quarks.
Polarisation studies are much more direct
for the top quark which is a short lived object and decays
before hadronisation takes place. \linebreak[4]
Angular distributions of
leptons from semileptonic decays of the top quark can provide
a very interesting information on the space-time structure of
the corresponding electroweak current \cite{jk5}.

First order QCD corrections to the energy spectra of charged
leptons in the decays of $c$ and $b$ quarks have been calculated
analytically in ref.\cite{ccm} and \cite{corbo}.
The results of a later analytical calculation
presented in ref.\cite{jk1}
disagree with those of \cite{ccm} and \cite{corbo}
and agree with a Monte Carlo for charm and bottom \cite{ali}
as well as with numerical results for top \cite{jk}.
In \cite{jk} the formulae  from \cite{ali}
were used and the accuracy achieved was good enough
to observe discrepancies with \cite{ccm} and \cite{corbo}.
The formulae for the virtual corrections used in \cite{ali} and
\cite{jk} were taken from the classic articles on muon decays
\cite{radcor}.

The formulae given in the present article agree with the ones
given in \cite{jk1} but are much simpler. They agree
also with those given in \cite{czajk91} for the joint angular
and energy charged lepton distribution in top quark decays.
In the latter case simplification is even more striking.
Some numerical results for polarised charm and bottom quarks have
been already presented in \cite{cjkk}.

In section \ref{Angene} the formulae are given for the QCD
corrected distributions of leptons originating from weak decays
of polarised heavy quarks:
subsection \ref{triple} contains our
main result -- the triple differential angular and energy
distributions of leptons,
and in subsections \ref{x-theta} and \ref{sec:1.4}
the double differential distributions are described.
The corresponding formulae for a massless quark in the final state
are presented in section \ref{Massless}. In section \ref{Angular}
QCD corrections to the angular distributions in charm and bottom
decays are presented. For the sake of completeness we summarize
our calculations based on \cite{jk1} and
\cite{czajk91} in Appendix A.

\section{Angular and energy distributions}
\label{Angene}
In this section we give the formulae
\footnote{Fortran77 version of the formulae given in this article
is available upon request from jezabek@chopin.ifj.edu.pl}
for the distributions of leptons from the weak decays
of a polarised heavy quark $Q$ of mass $m_1$ and the
weak isospin $I_3=\pm 1/2$. The mass of the quark $q$
originating from the decay is denoted by $m_2$, and
$\epsilon= m_2/m_1$.
These formulae are written in the $Q$ rest frame.
However, the two variables which we use, namely
the scaled energy
$x=2{\ell}^0/m_1=2 Q\cdot\ell/m_1^2$,
where $\ell$ is the four-momentum of the charged lepton,
and the scaled effective mass squared of the leptons
$y=(\ell+\nu)^2/m_1^2$
are Lorentz invariants.
For the massless lepton case
considered in the present article
also the third variable $\cos\theta$,
can be related to the scalar product $\ell\cdot s$,
where $s^\mu=(0,\vec s)$ is
the spin four-vector of the decaying quark and
$\theta$ denotes the angle between $\vec s$
and the direction of the charged lepton.
$S\equiv|\vec s|=1$ corresponds to fully polarised,
$S=0$ to unpolarised decaying quarks.

The distribution of the neutrino for $I_3=\pm 1/2$
is given by the formulae describing the distribution
of the charged lepton from the decay of $Q$ with the weak
isospin $I_3=\mp 1/2$.

\subsection{Triple differential distributions}
\label{triple}
The QCD corrected triple differential distribution
for the semileptonic decay  of the polarised quark
with $I_3=\pm 1/2$ can be written in the following way
\begin{eqnarray}
{{\rm d}\Gamma^{\pm} \over {\rm d}x\,{\rm d}y\,{\rm d}\cos\theta  }\,
=&&\, \left.
{G_{_F}^2 m_1^5\over32 \pi^3}\,
{|V_{CKM}|^2\over(1-y/\bar y)^2 +\gamma^2}\,
\right(
{\rm F}^\pm_0(x,y) + S\cos\theta\,{\rm J}^\pm_0(x,y)
\nonumber\\
&&\left.\qquad -{2\alpha_s\over3\pi}
  \left[ {\rm F}^\pm_1(x,y) + S\cos\theta\,{\rm J}^\pm_1(x,y)\right]
\right)
\label{eq:1}
\end{eqnarray}
$V_{CKM}$ denotes the element of the Cabbibo--Kobayashi--Maskawa
quark mixing matrix corresponding to the decay channel under
consideration.
The effects of $W$ propagator are included as a factor
$1/[(1-y/\bar y)^2+\gamma^2]$, where $\bar y= m_{\rm w}^2/m_1^2$ and
$\gamma=\Gamma_{\rm w}/m_{\rm w}$; c.f. \cite{jk1}.
These effects are important for the top quark \cite{jk1,jk4}.
For charm and bottom quarks $m_1$ is much smaller than
the mass of $W$ boson and the four-fermion limit can be employed
in which the factor mentioned above is replaced by $1$.

The Dalitz variables $x$ and $y$ fulfil kinematic constraints:
\begin{eqnarray}
0\le &x& \le x_m= 1-\epsilon^2
\nonumber\\
0\le &y& \le y_m= x(x_m-x)/(1-x)
\label{eq:2}
\end{eqnarray}
or
\begin{eqnarray}
0\le y \le  (1-\epsilon)^2 &&
\nonumber\\
w_-\le x \le w_+ &&
\label{eq:3}
\end{eqnarray}
The functions $w_\pm$ and other useful kinematic functions are
defined as follows:
\begin{eqnarray}
p_0 &=& (1-y+\epsilon^2)/2
\nonumber\\
p_3 &=& \sqrt{p_0^2 - \epsilon^2}
\nonumber\\
p_\pm &=& p_0 \pm p_3 = 1 - w_\mp
\nonumber\\
Y_p &=& {\textstyle{1\over 2}}\ln\left(p_+/ p_-\right)
=\ln\left(p_+/\epsilon\right)
\nonumber\\
Y_w &=& {\textstyle{1\over 2}}\ln\left(w_+/w_-\right)
=\ln\left(w_+/\sqrt{y}\right)
\nonumber\\
z_m &=& (1-x)(1-y/x)
\label{eq:4}
\end{eqnarray}
\pagebreak[4]

\noindent
The functions ${\rm F}^\pm_0(x,y)$ and ${\rm J}^\pm_0(x,y)$
corresponding to Born approximation read:
\begin{eqnarray}
{\rm F}^+_0(x,y) &=& x (x_m-x)
\label{eq:5}\\
{\rm J}^+_0(x,y) &=& {\rm F}^+_0(x,y)
\label{eq:6}\\
{\rm F}^-_0(x,y) &=& (x-y) (x_m-x+y)
\label{eq:7}\\
{\rm J}^-_0(x,y) &=& (x-y) (x_m-x+y-2y/x)
\label{eq:8}
\end{eqnarray}
The first order QCD corrections and the corresponding
functions ${\rm F}^\pm_1(x,y)$
have been calculated in \cite{jk1} whereas ${\rm J}^+_1(x,y)$
has been given in \cite{czajk91}.
In the course of the present work we rederived these old
results and from them we obtained
the much simpler expressions given
in this article. The result for ${\rm J}^-_1(x,y)$ is new.
The formulae for ${\rm F}^\pm_1(x,y)$ and ${\rm J}^+_1(x,y)$
read\footnote{Throughout this article
polylogarithms are defined as real functions. In particular:
$${\rm Li}_2(x) = -\int_0^x {\rm d}y\, {\ln |1-y|/  y}$$
$${\rm Li}_3(x) = \int_0^x {\rm d}y\, {{\rm Li}_2(y) / y}$$ }:
\begin{eqnarray}
{\rm F}^\pm_1(x,y) &=& \,{\rm F}^\pm_0(x,y)\, \Phi_0 \, +\,
\sum_{n=1}^5\, A^\pm_n\,\Phi_n\, +\, A^\pm_6
\label{F1pm}
\end{eqnarray}
\begin{eqnarray}
{\rm J}^\pm_1(x,y) &=& \,{\rm J}^\pm_0(x,y)\, \Phi_0 \, +\,
\sum_{n=1}^5\, B^\pm_n\Phi_n \, +\, B^\pm_6
\label{J1pm}
\end{eqnarray}
where
\begin{eqnarray}
\Phi_0 &=& {2p_0\over p_3}
\left[
 {\rm Li}_2\left(1-{1-x\over p_+}\right)
+ {\rm Li}_2\left(1-{1-y/x\over p_+}\right)
- {\rm Li}_2\left(1-{1-x\over p_-}\right)
\right.
\nonumber\\
&&\left.  \qquad
- {\rm Li}_2\left(1-{1-y/x\over p_-}\right)
+ {\rm Li}_2\left(w_-\right)
- {\rm Li}_2\left(w_+\right)
+ 4Y_p\ln\epsilon
\right]
\nonumber\\
&& + 4\left(1 -{p_0\over p_3} Y_p\right)\ln\left(z_m-\epsilon^2\right)
 - 4\ln z_m
\nonumber\\
\Phi_1 &=&
{\rm Li}_2\left(w_-\right)
+ {\rm Li}_2\left(w_+\right)
- {\rm Li}_2(x)
- {\rm Li}_2(y/x)
\nonumber\\
\Phi_2 &=& Y_p/p_3
\nonumber\\
\Phi_3 &=& {\textstyle{1\over2}}\ln\epsilon
\nonumber\\
\Phi_4 &=& {\textstyle{1\over2}}\ln(1-x)
\nonumber\\
\Phi_5 &=& {\textstyle{1\over2}}\ln(1-y/x)
\label{eq:11}
\end{eqnarray}
\pagebreak[4]
\begin{eqnarray}
A^+_1 &=& x + \epsilon^2x
\nonumber\\
A^+_2 &=&
           - p_3^2 \, (  3 + 2x + y   +  \epsilon^2  )
           - 2 \epsilon^2  y
\nonumber\\
A^+_3 &=&
           -(1-y)( 3 + 2x +y )
           + 2\epsilon^2  (  1 + 5x )
           + \epsilon^4
\nonumber\\
A^+_4 &=&
           5 - 5x
           - \epsilon^2  ( 4 + 5x )
           - \epsilon^4
\nonumber\\
A^+_5 &=&
            - 2xy    + 9x - 4x^2 - 2y - y^2
           - 7 \epsilon^2  x
\nonumber\\
A^+_6 &=&
             ( - 4xy  + 4y - y^2 + y^2/x )/2
           + \epsilon^2  [ x  + y   - x/(1-y/x)] /2
\label{eq:12}\\
\nonumber\\
B^+_1 &=&  - x + \epsilon^2  x
\nonumber\\
B^+_2 &=&
            p_3^2\,  [\,   ( 5 - 2x - y - 2y/x - 2/x )
           + \epsilon^2  ( - 1 + 2/x )  \,   ]  \,
           + \epsilon^2  ( - xy - x
\nonumber\\
&&\         + 2y - y/x - y^2/x )
           + \epsilon^4  (  x   + y/x )
\nonumber\\
B^+_3 &=&
            (1-y)  ( 5 - 2x - y - 2y/x - 2/x )
           + 2 \epsilon^2  ( - 3  + 7x  + 2/x  )
\nonumber\\
&&         + \epsilon^4  ( 1 - 2/x  )
\nonumber\\
B^+_4 &=&
            - 3 + x + 2/x
           + \epsilon^2  ( 4 - 9x - 4/x )
           + \epsilon^4  ( - 1 + 2/x  )
\nonumber\\
B^+_5 &=&
              - 2xy + 3x - 4x^2 + 6y - y^2 - 2y^2/x
           - 7  \epsilon^2  x
\nonumber\\
B^+_6 &=&
           ( 2 - 2x^2 + 2y - 3y^2 - 2y/x + 3y^2/x )/2
           + \epsilon^2  [ - 4 + x + y  + 2y/x
\nonumber\\
&&\           - x/(1-y/x)] /2
           + \epsilon^4
\label{eq:13}\\
 && \nonumber\\
A^-_1 &=&
           - 2xy  + x  + y  + y^2
           + \epsilon^2  ( x - y )
\nonumber\\
A^-_2 &=&
            p_3^2 \, ( - 5  + 2x  - 3y + \epsilon^2 )
           - 2  \epsilon^2  y
\nonumber\\
A^-_3 &=&
            (1-y)  ( - 5 + 2x - 3y )
           + 6 \epsilon^2  ( 1  + x  - 2y  )
           - \epsilon^4
\nonumber\\
A^-_4 &=&
            5 + 4xy - 5x + 3y - 5y^2 - 2y^2/x
           + \epsilon^2  ( - 4 - 5x + 11y )
           - \epsilon^4
\nonumber\\
A^-_5 &=&
                6xy + 9x - 4x^2 - 11y - 2y^2 + 2y^2/x
                + 7 \epsilon^2  ( - x + y )
\nonumber\\
A^-_6 &=&
              ( 3xy + 2y - 3y^2 - 2y^2/x )/2
           + \epsilon^2   [ 3y - y/(1-x)] /2
\label{eq:14}\\
\nonumber\\
B^-_1 &=&
 - 2xy - x  - 5y + y^2 -2y^2/x
           + \epsilon^2  ( x - y )
\nonumber\\
B^-_2 &=&
            p_3^2 \, [\,  ( 3 + 10x + y + 10y/x - 2/x )
           + \epsilon^2  (  1 + 2/x )\,    ]
           + \epsilon^2  ( xy + x - 2y
\nonumber\\
&&\         + y/x + y^2/x )
           - \epsilon^4  ( x + y/x )
\nonumber\\
B^-_3 &=&
          (1-y)  ( 3 + 10x + y + 10y/x - 2/x )
           + 2 \epsilon^2  ( - 1 + 5x - 4y
           - 2y/x
\nonumber\\
&&\       + 2/x )
           - \epsilon^4  ( 1 + 2/x )
\nonumber\\
B^-_4 &=&
            - 3 + 12xy + x - 7y - y^2 - 12y/x + 8y^2/x + 2/x
             + \epsilon^2  (  4 - 9x
\nonumber\\
&&\            + 7y - 4/x )
           + \epsilon^4  ( - 1  + 2/x )
\nonumber\\
B^-_5 &=&
               6xy - 9x - 4x^2 - y - 2y^2 + 10y^2/x
           + 7  \epsilon^2  ( - x + y   )
\nonumber\\
B^-_6 &=&
           ( 2  - 5xy - 2x^2 + 2y + 7y^2 - 2y/x - 2y^2/x )/2
           + \epsilon^2  [ - 4  - 9y + 2y/x
           \nonumber\\
&&\         + y/(1-x)] /2
           + \epsilon^4
\label{eq:15}
\end{eqnarray}

\subsection{\protect\boldmath Double differential $x-\theta$
distributions}
\label{x-theta}
For the top quark in the narrow $W$ width limit $\gamma\to 0$
the mass squared of the leptons is fixed at value
$y = (m_{\rm w}/m_t)^2$. Thus, in this
limit the triple differential distributions of subsection
\ref{triple} reduce to the double differential
$x-\theta$ distributions. In the four-fermion limit
($m_{\rm w}\to\infty$) the double differential
$x-\theta$ distributions can be calculated from the following
formulae:
\begin{eqnarray}
\left.
{{\rm d}\Gamma^{\pm} \over {\rm d}x\,{\rm d}\cos\theta  }\;
=\;  \int^{y_m}_0\,{\rm d}y\;
{{\rm d}\Gamma^{\pm} \over {\rm d}x\,{\rm d}y\,{\rm d}\cos\theta  }\,
=\,
{G_{_F}^2 m_1^5\over32 \pi^3}\,|V_{CKM}|^2\,
\right(
f^\pm_0(x)\qquad
\nonumber\\
\left.
+ S\cos\theta\, j^\pm_0(x)
\; -{2\alpha_s\over3\pi}
  \left[ f^\pm_1(x) + S\cos\theta\, j^\pm_1(x)\right]
\right)
\label{eq:1.3.1}
\end{eqnarray}
where
\begin{eqnarray}
f^+_0(x) &=& x^2 (x_m-x)^2/(1-x)
\label{eq:1.3.2}\\
j^+_0(x) &=& f^+_0(x,y)
\label{eq:1.3.3}\\
f^-_0(x) &=& {\textstyle{1\over6}} x^2 (x_m-x)^2
[\, 3 - 2x + \epsilon^2 + 2\epsilon^2/(1-x)\, ]/(1-x)^2
\label{eq:1.3.4}\\
j^-_0(x) &=& {\textstyle{1\over6}} x^2 (x_m-x)^2
[\, 1 - 2x + \epsilon^2 - 2\epsilon^2/(1-x)\, ]/(1-x)^2
\label{eq:1.3.5}
\end{eqnarray}
and
\begin{eqnarray}
f^\pm_1(x) &=&
\int^{y_m}_0 {\rm d}y\, {\rm F}^\pm_1(x,y)
\label{eq:1.3.6}\\
j^\pm_1(x) &=&
\int^{y_m}_0{\rm d}y\, {\rm J}^\pm_1(x,y)
\label{eq:1.3.7}
\end{eqnarray}
The integrals in eqs.(\ref{eq:1.3.6})-(\ref{eq:1.3.7})
can be calculated numerically.
In the massless limit $\epsilon\to 0$ analytic formulae for
$f^\pm_1(x)$  and $f^\pm_1(x)$ have been derived. These formulae
are listed in subsection \ref{xdis0}.
The results for charm and bottom quarks
have been already published in \cite{cjkk}.
For the polarisation independent parts of the distributions they
are in perfect agreement with the results of \cite{jk1}.
Thus, they are in conflict with the results of \cite{ccm}
and \cite{corbo}; see section 5 in \cite{jk1}
where a detailed comparison
is given and cross checks descibed.\\
A convenient way to present the results is to
express the double differential distributions as  products
of the energy distributions and the asymmetry functions, c.f.
eqs.(5)-(7) in \cite{cjkk},
\begin{eqnarray}
{{\rm d}\Gamma^{\pm} \over {\rm d}x\, {\rm d}\cos\theta  }\,=\,
{{\rm d}\Gamma^{\pm} \over {\rm d}x}\,
\left[\,1\, \pm\, \alpha_\pm(x) \cos\theta\,\right] \,/2
\label{eq:1.3.8}
\end{eqnarray}
where
\begin{equation}
\alpha_\pm(x) = \pm\, S\,\left[\,
j^\pm_0(x) - {\textstyle{2\alpha_s\over3\pi}}
j^\pm_1(x)\, \right]\, /\,
\left[\,  f^\pm_0(x) - {\textstyle{2\alpha_s\over3\pi}}
f^\pm_1(x) \right]
\label{eq:1.3.9}
\end{equation}
The functions $\alpha_\pm(x)$ for charm and bottom quarks
have been given in $\cite{cjkk}$.

\subsection{\protect\boldmath Double differential $y-\theta$
distributions}
\label{sec:1.4}
We follow the normalisation convention of \cite{jk0} and write
the double differential $y-\theta$ distributions
in the following way:
\begin{eqnarray}
{{\rm d}\Gamma^{\pm} \over {\rm d}y\,{\rm d}\cos\theta  }\;
=\;  \int^{w_+}_{w_-}\,{\rm d}x\;
{{\rm d}\Gamma^{\pm} \over {\rm d}x\,{\rm d}y\,{\rm d}\cos\theta  }\,
=\,
{G_{_F}^2 m_1^5\over192 \pi^3}\,
{|V_{CKM}|^2\over(1-y/\bar y)^2 +\gamma^2}\,
\nonumber\\
\times {1\over 2}\, \left(
{\cal F}_0(y) +
 S\cos\theta\, {\cal J}^\pm_0(y)
\,  -{2\alpha_s\over3\pi}
  \left[ {\cal F}_1(y) + S\cos\theta\, {\cal J}^\pm_1(y)\right]\,
\right)
\label{eq:1.4.1}
\end{eqnarray}
where
\begin{eqnarray}
{\cal F}_0(y) &=& 4 p_3\,
\left[ \, (1-\epsilon^2)^2 + y(1+\epsilon^2) - 2 y^2\, \right]
\label{eq:1.4.3}\\
{\cal J}_0^+(y) &=& {\cal F}_0(y)
\label{eq:1.4.4}\\
{\cal J}_0^-(y) &=& {\cal F}_0(y)  +48 y \left(y Y_w - p_3\right)
\label{eq:1.4.5}
\end{eqnarray}
and
\begin{eqnarray}
{\cal F}_1(y) &=&
12 \int^{w_+}_{w_-} {\rm d}x\, {\rm F}^\pm_1(x,y)
\label{eq:1.4.6}\\
{\cal J}^\pm_1(y) &=&
12 \int^{w_+}_{w_-} {\rm d}x\,  {\rm J}^\pm_1(x,y)
\label{eq:1.4.7}
\end{eqnarray}
Eq.(\ref{eq:1.4.6}) provides a very non-trivial cross check
of this calculation. It has been verified both analytically
and numerically that for  ${\rm F}^\pm_1(x,y)$ given by
eq.(\ref{F1pm}) the integrals in (\ref{eq:1.4.6})  are equal.
Moreover, ${\cal F}_1(y)$ can be calculated in an easier way,
see formula (2.6) in \cite{jk0}, if the phase space integrals
over the momenta of leptons are performed first.
This cross check is also fulfilled. It has been shown,
see sect.5 of \cite{jk1}, that when
${\cal F}_1(y)$ is analytically continued
it agrees with the analytic results of \cite{CGvN}
for QCD corrections to the width of $W$ boson.
All these tests indicate that our formulae for ${\rm F}^\pm_1(x,y)$
are correct. Since all the calculations for both
${\rm F}^\pm_1(x,y)$ and ${\rm J}^\pm_1(x,y)$ were performed
in parallel using the same algebraic computer programs we believe
that the same is true also for ${\rm J}^\pm_1(x,y)$.\\
Analytic expression for ${\cal F}_1(y)$ has been first given
in \cite{jk0}. For the sake of completeness we rewrite this
expression using the notation of the present article and some
identities for dilogarithms.  Our formulae for ${\cal F}_1(y)$
and ${\cal J}^\pm_1(y)$ read:
\begin{eqnarray}
{\cal F}_1^{ }(y) \; &=& \;
{\cal F}_0(y)\Psi_0 \,  +\, {\cal A}_1 \Psi_1\,
+\, {\cal A}_2 Y_w \, +\, {\cal A}_3 Y_p \,
+\, {\cal A}_4 p_3\ln\epsilon\, +\, {\cal A}_5 p_3
\label{eq:F1y}\\
{\cal J}_1^+(y) &=& {\cal J}_0^+(y)\, \Psi_0   \,
  + \, {\cal B}^+_1 \Psi_1\,
+\, {\cal B}^+_2 Y_w \, +\, {\cal B}^+_3 Y_p \,
+\, {\cal B}^+_4 p_3\ln\epsilon\, +\, {\cal B}^+_5 p_3
\nonumber\\
&&\qquad +\, {\cal B}^+_6 Y_w Y_p \, +\, {\cal B}^+_7 Y_w \ln\epsilon
\label{eq:J1PY}\\
{\cal J}_1^-(y)\; &=&\; {\cal J}_0^-(y)\, \Psi_0   \,
 + \, {\cal B}^-_1 \Psi_1\,
+\, {\cal B}^-_2 Y_w \, +\, {\cal B}^-_3 Y_p \,
+\, {\cal B}^-_4 p_3\ln\epsilon\, +\, {\cal B}^-_5 p_3
\nonumber\\
&&\qquad +\, {\cal B}^-_6 Y_w Y_p \, +\, {\cal B}^-_7 Y_w \ln\epsilon
\, +\, 48 y^2\, \left({\cal B}^-_8 \, +\, {\cal B}^-_9 p_0/p_3 \right)
\label{eq:J1MY}
\end{eqnarray}
where
\begin{eqnarray}
\Psi_0 &=&  2\, \left[\, 4{\rm Li}_2(2p_3/p_+) - 4Y_p \ln(2p_3/p_+)
- \ln p_-\ln w_+ +  \ln p_+\ln w_- \,\right] p_0/p_3
\nonumber\\
&&\quad + 8\ln\left(2p_3\right) - 2\ln y
\label{eq:Psi0}\\
\Psi_1 &=& 4\, {\rm Li}_2(w_-) \;-\;  4\,{\rm Li}_2(w_+)
\label{eq:Psi1}
\end{eqnarray}
and
\begin{eqnarray}
{\cal A}_1 &=&  {\cal F}_0(y)\, p_0/p_3
\nonumber\\
{\cal A}_2 &=&
- 8 (1-\epsilon^2) \left[ 1 +y -4 y^2-
\epsilon^2 (2-y) +\epsilon^4 \right]
\nonumber\\
{\cal A}_3 &=&
- 2 \left[ 3 + 6 y   -21 y^2  + 12 y^3
-\epsilon^2 (1+12y+5y^2) + \epsilon^4(11+2y)- \epsilon^6 \right]
\nonumber\\
{\cal A}_4 &=&
- 12 \left[ 1  + 3 y  - 4 y^2
-\epsilon^2 (4-y) + 3 \epsilon^4 \right]
\nonumber\\
{\cal A}_5 &=&
- 2 \left[ 5+9y-6y^2 -\epsilon^2( 22 - 9y) + 5 \epsilon^4 \right]
\\
\nonumber\\
{\cal B}^+_1 &=&  5 - 9y^2 + 4y^3  - \epsilon^2 (8-8y+6y^2)
 +\epsilon^4+2 \epsilon^6
\nonumber\\
{\cal B}^+_2 &=&
   -8 \left[ 1+4 y-y^2-\epsilon^2 (3+3 y-4 y^2)
   +\epsilon^4 (3-y) - \epsilon^6 \right]
\nonumber\\
{\cal B}^+_3 &=&
- 2 \left[ 3 (1-y)^2 (1+4 y) + \epsilon^2 (11-5y^2)
-\epsilon^4(13-2y)-\epsilon^6 \right]
\nonumber\\
{\cal B}^+_4 &=&
-12 \left[ (1-y) (1+4 y) -\epsilon^2 (4-y)+3 \epsilon^4 \right]
\nonumber\\
{\cal B}^+_5 &=&
   2\, \left[15-y+2 y^2-\epsilon^2 (12+7 y)-3 \epsilon^4\right]
\nonumber\\
{\cal B}^+_6 &=&
 -24 \left( 1+y-\epsilon^2 \right)\left( 2p_3+y \epsilon^2/p_3 \right)
\nonumber\\
{\cal B}^+_7 &=&
- 24 \left(\, 1-y^2-2\epsilon^2 + \epsilon^4\, \right)
\\
\nonumber\\
{\cal B}^-_1 &=&
   5-42y+45y^2+4y^3 -2\epsilon^2 (4+8y+3y^2)
              + \epsilon^4 + 2\epsilon^6
\nonumber\\
{\cal B}^-_2 &=&
        8 \left[ -1+8y+10y^2
     + \epsilon^2(3-9y-4y^2)- \epsilon^4(3-y)+ \epsilon^6\right]
\nonumber\\
{\cal B}^-_3 &=&
        2 \left[ - 3(1-y)(3 - 23y - 4y^2 )
          + \epsilon^2(1+24y+5y^2)  +\epsilon^4(7-2y)
     + \epsilon^6 \right]
\nonumber\\
{\cal B}^-_4 &=&  - 12  \left[\, 3 - 23y - 4y^2
     - \epsilon^2(6-y) + 3\epsilon^4\,  \right]
\nonumber\\
{\cal B}^-_5 &=&
       - 2\, \left[\, -15 + 37y - 2y^2
     + \epsilon^2(12+7y) + 3\epsilon^4\,\right]
\nonumber\\
{\cal B}^-_6 &=&
       - 24\, \left[\,  2p_3 (1 - 5y - \epsilon^2)
     \,-\, \epsilon^2(1 + y -\epsilon^2)y/p_3 \, \right]
\nonumber\\
{\cal B}^-_7 &=&
       - 24\, \left[ (1-y)(1-5y)
     - 2\epsilon^2(1-y) + \epsilon^4 \right]
\nonumber\\
{\cal B}^-_8 &=&
    4\, \left[ \,
    {\rm Li}_2\left( {\textstyle {w_-p_-\over w_+p_+}} \right)
      - {\rm Li}_2\left( {p_-/ p_+} \right)  \, \right]
      +  {\rm Li}_3(w_+) - {\rm Li}_3(w_-)
      + 8 Y_p\ln(w_+)
\nonumber\\
  && - Y_w\,\left[\, {\rm Li}_2(w_-)+{\rm Li}_2(w_+) + 8\ln(2p_3)
      - 2\ln y -4\ln(w_+) \, \right]
\nonumber\\
{\cal B}^-_9 &=&
        4\,\left[   {\rm Li}_3(1)
        +  {\rm Li}_3\left( {\textstyle {w_-p_-\over w_+p_+}} \right)
        -  {\rm Li}_3({p_-/ p_+}) -  {\rm Li}_3(w_-/w_+) \right]
\nonumber\\
&&        + 2Y_w\,\left[\,  {\rm Li}_2(w_+) - {\rm Li}_2(w_-)
        - 4{\rm Li}_2(2p_3/p_+) + 4Y_p\ln(2p_3/p_+)
        \right.
\nonumber\\
&&\quad   \left.     + 2Y_w\ln p_+    - 4Y_p\ln w_+
       \,  \right]
    \;+\; 4 Y_p\, \left[\,
    {\rm Li}_2\left( {\textstyle {w_-p_-\over w_+p_+}} \right)
        - {\rm Li}_2({p_-/ p_+})\, \right]
\end{eqnarray}

\section{Massless case}
\label{Massless}
\subsection{Triple differential distributions}
For $m_2=0$ the functions in eq.(\ref{eq:1}) simplify considerably.
The limit $\epsilon\to 0$ can be easily obtained for
${\rm F}^\pm_1(x,y)$ and ${\rm J}^\pm_1(x,y)$.
In this limit
\begin{eqnarray}
{\rm F}^+_0(x,y) &=& x (1-x)
\label{eq:5a}\\
{\rm J}^+_0(x,y) &=& {\rm F}^+_0(x,y)
\label{eq:6a}\\
{\rm F}^-_0(x,y) &=& (x-y) (1-x+y)
\label{eq:7a}\\
{\rm J}^-_0(x,y) &=& (x-y) (1-x+y-2y/x)
\label{eq:8a}
\end{eqnarray}
and
\begin{eqnarray}
{\rm F}^+_1(x,y) &=&
   {\rm F}^+_0\,\Phi_0 + x\Phi_1 - ( 3 + 2x +y )\Phi_{2\diamond 3}
   + 5(1-x)\Phi_4 + ( - 2xy
\nonumber\\
&&\   + 9x     - 4x^2 - 2y - y^2)\Phi_5
      + y  (  4 - 4x - y + y/x )/2
\label{eq:16}\\
\nonumber\\
{\rm F}^-_1(x,y) &=&
        {\rm F}^-_0\,\Phi_0 + ( - 2xy  + x  + y  + y^2)\Phi_1
        + ( - 5  + 2x  - 3y )\Phi_{2\diamond 3}
\nonumber\\
&&
        + (  5 + 4xy - 5x + 3y - 5y^2 - 2y^2/x)\Phi_4
       + (  6xy + 9x  - 4x^2
\nonumber\\
&&\       - 11y   - 2y^2 + 2y^2/x )\Phi_5
        + y  (  2 + 3x - 3y - 2y/x )/2
\label{eq:17}\\
\nonumber\\
{\rm J}^+_1(x,y) &=&
       {\rm J}^+_0\,\Phi_0 - x\Phi_1
       +  ( 5 - 2x - y - 2y/x - 2/x )\Phi_{2\diamond 3}
       +  ( - 3 + x
\nonumber\\
&&\        + 2/x)\Phi_4
       +  ( - 2xy + 3x - 4x^2 + 6y - y^2 - 2y^2/x)\Phi_5
\nonumber\\
&&       +  ( 2 - 2x^2 + 2y - 3y^2 - 2y/x + 3y^2/x )/2
\label{eq:18}\\
\nonumber\\
{\rm J}^-_1(x,y) &=&
      {\rm J}^-_0\, \Phi_0
      + ( - 2xy - x  - 5y + y^2 -2y^2/x )\Phi_1
      + ( 3 + 10x + y
\nonumber\\
&&\
      + 10y/x - 2/x )\Phi_{2\diamond 3}
      + ( - 3 + 12xy + x - 7y - y^2 - 12y/x
\nonumber\\
&&\
+ 8y^2/x + 2/x )\Phi_4
+ ( 6xy - 9x - 4x^2 - y - 2y^2 + 10y^2/x)\Phi_5
\nonumber\\
&&         +  ( 2  - 5xy - 2x^2 + 2y + 7y^2 - 2y/x - 2y^2/x )/2
\label{eq:19}
\end{eqnarray}
where
\begin{eqnarray}
&\Phi_0\ \ &=\ {\textstyle {\pi^2\over 3}} + 2{\rm Li}_2(x)
 + 2{\rm Li}_2(y/x) + \ln^2\left({\textstyle {1-y/x\over 1-x}}\right)
\nonumber\\
&\Phi_1\ \ &=\  {\textstyle {\pi^2\over 6}} + {\rm Li}_2(y)
           - {\rm Li}_2(x)   - {\rm Li}_2(y/x)
\nonumber\\
&\Phi_{2\diamond 3} &=\ {\textstyle{1\over2}} (1-y)\ln(1-y)
\nonumber\\
&\Phi_4\ \ &=\ {\textstyle{1\over2}}\ln(1-x)
\nonumber\\
&\Phi_5\ \ &=\ {\textstyle{1\over2}}\ln(1-y/x)
\end{eqnarray}
Eqs.(\ref{eq:16})-(\ref{eq:18})
agree with those given in the literature; c.f. eqs.(3.9) and (4.9)
in \cite{jk1} and eq.(3.3) in \cite{czajk91}.

\subsection{\protect\boldmath Massless limit for $x-\theta$
distributions}
\label{xdis0}
For $\epsilon=0$ explicit analytic formulae can be derived
for the double differential $x-\theta$ distributions.
The functions
$f^\pm_0(x)$, $j^\pm_0(x)$,
$f^\pm_1(x)$  and $j^\pm_1(x)$  which appear in eq.(\ref{eq:1.3.1})
are given by the following expressions,
cf. eqs.(3.10) and (4.10) in \cite{jk1}:
\begin{eqnarray}
f_0^+(x) &=& x^2(1-x)
\label{eq:1.3.10}\\
f_0^-(x) &=& x^2(3-2x)/6
\label{eq:1.3.11}\\
j_0^+(x) &=& f_0^+(x)
\label{eq:1.3.12}\\
j_0^-(x) &=& x^2(1-2x)/6
\label{eq:1.3.13}\\
f_1^+(x) &=&  f_0^+(x) \phi_0(x)
       + (1-x)\left[
                {\textstyle{1\over 6}} (5+8x+8x^2)\ln(1-x) \right.
\nonumber\\
&&\  \left.          +{\textstyle{1\over12}} x(10-19x)  \right]
\label{eq:1.3.14}\\
f_1^-(x) &=&  f_0^-(x) \phi_0(x)
       + {\textstyle{1\over36}}
              (41-36x+42x^2-16x^3)\ln(1-x)
\nonumber\\
&&\           +{\textstyle{1\over72}} x(82-153x+86x^2)
\label{eq:1.3.15}\\
j_1^+(x) &=&  j_0^+(x) \phi_0(x)
       + (1-x)\left[
           {\textstyle{1\over 6}}( - 3 + 12x + 8x^2 + 4/x)\ln(1-x)
         \right.
\nonumber\\
&&\ \left.  + {\textstyle{1\over12}} ( 8 - 2x - 15x^2 ) \right]
\label{eq:1.3.16}\\
j_1^-(x) &=&  j_0^-(x) \phi_0(x)
       + {\textstyle{1\over36}}
           ( 11 - 36x + 14x^2 - 16x^3 - 4/x )\ln(1-x)
\nonumber\\
&&\       + {\textstyle{1\over72}}
             ( - 8 + 18x - 103x^2 + 78x^3 )
\label{eq:1.3.17}\\
\phi_0(x) &=& 2{\rm Li}_2(x) + 2 \pi^2/3 + \ln^2(1-x)\qquad
\end{eqnarray}

\subsection{\protect\boldmath Massless limit for $y-\theta$
distributions}
\label{ydis0}
For $\epsilon\to 0$ the functions in eq.(\ref{eq:1.4.1}) read,
cf. eqs.(3.1)-(3.2) in \cite{jk0} :
\begin{eqnarray}
\lefteqn{
{\cal F}_0(y) \; =\; 2 (1-y)^2 (1+2y) }
\label{eq:2.3.1}    \\
\lefteqn{
{\cal J}_0^+(y) \; =\; {\cal F}_0(y) }
\label{eq:2.3.2}    \\
\lefteqn{
{\cal J}_0^-(y) \; =\; 2 (1-y)(1 -11y-2y^2) - 24y^2\ln y   }
\label{eq:2.3.3}   \\
\lefteqn{
{\cal F}_1(y) \; =\;
{\cal F}_0\,
\left[\, {\textstyle {2\over3}}\pi^2+4{\rm Li}_2(y)+2\ln y\ln(1-y)\,\right]
      - (1-y)(5+9y- 6y^2) }
\nonumber\\  &&
      +\, 4y(1-y-2y^2)\ln y
      + 2(1-y)^2(5+4y)\ln(1-y)
\label{eq:F1y0}\\
\lefteqn{
{\cal J}_1^+(y) \; =\;
      - {\textstyle {2\over3}}\pi^2(1-y)(1+y+4y^2)
      + (1-y)(15-y+2y^2)   }
\nonumber\\   &&
      +\, 4(1-y)(5+5y-4y^2)\,{\rm Li}_2(y)
      + 8(1-y)(2+2y-y^2)\ln(1-y)\ln y
\nonumber\\   &&
      +\, 8y(2+y-y^2)\ln y
      + 2(1-y)^2(5+4y)\ln(1-y)
\label{eq:J1Py0}\\
\lefteqn{
{\cal J}_1^-(y) \; = \;
      - {\textstyle {2\over3}}\pi^2(1+6y+9y^2-4y^3)
      + (1-y)(15-37y+2y^2)    }
\nonumber\\   &&
      +\, 240y^2\,\left[\, {\rm Li}_3(1) - {\rm Li}_3(y)\, \right]
  + 4\left[\, 5-42y+45y^2+4y^3 +18y^2\ln y\,\right]\,{\rm Li}_2(y)
\nonumber\\   &&
      +\, 8(1-y)(2-13y-y^2)\ln(1 - y)\ln y
      + 4y\,\left[\, 4-(19+\pi^2)y-2y^2 \, \right]\,\ln y
\nonumber\\   &&
      -\, 2(1-y)(1+19y+4y^2)\ln(1 - y)
\label{eq:J1My0}
\end{eqnarray}
\section{Angular distributions}
\label{Angular}
Angular distributions are related to the double differential
$y-\theta$ distributions discussed in subsection \ref{sec:1.4}
\begin{eqnarray}
{{\rm d}\Gamma^{\pm} \over {\rm d}\cos\theta  }\;
=\;  \int^{(1-\epsilon)^2}_{0}\,{\rm d}y\;
{{\rm d}\Gamma^{\pm} \over {\rm d}y\,{\rm d}\cos\theta  }
\label{eq:3.1}
\end{eqnarray}
with ${{\rm d}\Gamma^{\pm}/ {\rm d}y\,{\rm d}\cos\theta  }$
given in eq.(\ref{eq:1.4.1}).\\
For the top quark and the narrow $W$ width the double
differential distributions are proportional to Dirac delta
function $\delta\left( y-m_{\rm w}^2/m_t^2 \right)$ and the
integral in (\ref{eq:3.1}) is trivial. For the charm and for
the bottom quarks the four-fermion approximation can be used.
\begin{figure}[h]
\label{fig1}
\begin{center}
\leavevmode
\epsffile[70 320 550 500]{fig1.ps}
\caption{a) QCD reduction of the total decay rate: the ratios
${\cal R}(\epsilon)/{\cal R}_0(\epsilon)$  as functions of
$\epsilon$ for $\alpha_s\,=\, 0.1,\, 0.2,\, 0.3$
and $0.4$ -- solid, dashed, dotted and dash-dotted lines;
b) QCD correction to the angular asymmetry $\alpha^-(\epsilon)$ for
$\alpha_s\,=\, 0,\, 0.2$ and $0.4$ -- solid, dashed
and dotted lines, respectively.}
\end{center}
\end{figure}

In order to illustrate the size of the radiative corrections
discussed in the present article we employ the four-fermion
limit and write the resulting angular distributions in the
following way:
\begin{equation}
{{\rm d}\Gamma^{\pm} \over {\rm d}\cos\theta  }\;
=\; {\Gamma_0}\,{\cal R}(\epsilon)\,
\left[\, 1\; \pm \; {\alpha}^\pm(\epsilon)\,\cos\theta\, \right]
\label{eq:3.2}
\end{equation}
where $\Gamma_0 = G_{_F}^2 m_1^5|V_{CKM}|^2/384 \pi^3$.
In Born approximation for completely polarised quarks ($S=1$)
\begin{eqnarray}
{\cal R}_0(\epsilon) = 1 -8\epsilon^2+ 8\epsilon^6-\epsilon^8
-24\epsilon^4\ln\epsilon &&
\label{eq:3.3}\\
\alpha^+_0(\epsilon) = 1 &&
\label{eq:3.4}\\
\alpha^-_0(\epsilon) = {\textstyle{1\over 3}}\,
\left[\, 1 -12\epsilon^2-36\epsilon^4+44\epsilon^6+3\epsilon^8
\,-\, 24\epsilon^4\,(3+2\epsilon^2)\ln\epsilon \right]\,/\,
{\cal R}_0(\epsilon) &&
\end{eqnarray}

QCD corrections decrease the total decay rate. In Fig.1a the ratios
${\cal R}(\epsilon)/{\cal R}_0(\epsilon)$ are shown as functions of
$\epsilon$ for $\alpha_s\,=\, 0.1,\, 0.2,\, 0.3$
and $0.4$ as the solid, dashed,
dotted and dash-dotted lines, respectively. The corrections to
eq.(\ref{eq:3.4}) are of the order of one percent or smaller.
In Fig.1b the function $\alpha^-(\epsilon)$ is shown for
$\alpha_s\,=\, 0,\, 0.2$ and $0.4$ as the solid, dashed
and dotted lines, respectively.

\appendix
\section{Derivations and formulae}
In this Appendix we sketch our derivation of eq.(\ref{eq:1}).
The method has been described in our earlier papers
\cite{jk1} and \cite{czajk91}.
Many formulae presented here have been already given
in these articles.
We rewrite them using the notation adopted in the present article.
Some formulae are simplified and misprints corrected.
In the following $Q$, $q$, $\ell$, $\nu$ and $G$
denote the scaled four-momenta
of the decaying quark and the decay products:
quark, charged lepton, neutrino and gluon. The common scaling factor
is $1/m_1$, and $Q^2=1$, $q^2=\epsilon^2$, $\ell^2=\nu^2=G^2=0$.
Thus, the gluon is massless. However, at intermediate steps
we introduce the scaled gluon mass $\lambda_G$ in order
to regularize those expressions which are infrared divergent.
We introduce also $P=q+G$, $z=P^2$,
$s$ -- the spin four-vector
of the decaying quark and $R = Q - s$.
Some formulae are given only for
the weak isospin $I_3= {1/2}$ of the decaying quark.
This implies that the analogous formulae
for $I_3= -{1/2}$ can be obtained by
replacing $\ell$ with $\nu$ and vice versa.

\subsection{Differential decay rate}
The QCD corrected differential rate is given by the following formula
\begin{equation}
{\rm d}\Gamma^\pm = {\rm d}\Gamma_0^\pm + {\rm d}\Gamma_{1,3}^\pm
+ {\rm d}\Gamma_{1,4}^\pm
\label{eq:A1}
\end{equation}
where
\begin{equation}
{\rm d}\Gamma_0^\pm =  G_{_F}^2 m_1^5\,|V_{CKM}|^2\,
{\cal M}_{0,3}^\pm {\rm d}{\cal R}_3(Q;q,\ell,\nu) /\pi^5
\label{eq:A2}
\end{equation}
is Born approximation,
\begin{equation}
{\rm d}\Gamma_{1,3}^\pm = {\textstyle{2\over3}}\,
\alpha_s\, G_{_F}^2 m_1^5\,|V_{CKM}|^2\,
{\cal M}_{1,3}^\pm\, {\rm d}{\cal R}_3(Q;q,\ell,\nu) /\pi^6
\label{eq:A3}
\end{equation}
describes the virtual gluon contribution and
\begin{equation}
{\rm d}\Gamma_{1,4}^\pm = {\textstyle{2\over3}}\,
\alpha_s\,G_{_F}^2 m_1^5\,|V_{CKM}|^2\,
{\cal M}_{1,4}^\pm {\rm d}{\cal R}_4(Q;q,G,\ell,\nu) /\pi^7
\label{eq:A4}
\end{equation}
comes from real gluon emission.
Lorentz invariant n-body
phase space is defined as
\begin{equation}
{\rm d}{\cal R}_n(P;p_1,p_2,\dots,p_n) = \delta^{(4)}(P-{\bf\Sigma} p_i)
\prod_i {{\rm d}^3 {\bf p}_i\over 2p_{0,i}}
\label{eq:A5}
\end{equation}

\subsection{Three-body contributions}
In Born approximation the rates for the decays into three-body
final states  are proportional to the expressions
\begin{eqnarray}
{\cal M}_{0,3}^+ &=& q\cdot\nu\; R\cdot\ell \; /\;
\left[\,(1-y/\bar y)^2 + \gamma^2\,\right]
\nonumber\\
{\cal M}_{0,3}^- &=& q\cdot\ell\; R\cdot\nu \;/\;
\left[\,(1-y/\bar y)^2 + \gamma^2\,\right]
\label{eq:A6}
\end{eqnarray}
The formulae for ${\rm F}^\pm_0(x,y)$ and ${\rm J}^\pm_0(x,y)$
can be easily derived from (\ref{eq:A2}) and (\ref{eq:A6})
when Dalitz parametrization of the three-body phase space
\begin{equation}
{\rm d}{\cal R}_3(Q;q,\ell,\nu) = {\textstyle{1\over 32}}
{\rm d}x {\rm d}y {\rm d}\alpha {\rm d}(\cos\theta) {\rm d}\beta
\label{eq:A7}
\end{equation}
is employed. Integration over Euler angle $\beta$ is trivial,
and non-trivial integrals over $\alpha$ read, see Appendix B
in \cite{czajk91},
\begin{eqnarray}
\int{\rm d}\alpha\;s\cdot q &=& 2\pi S\cos\theta
[ (1+y-q^2)/2 - y/x ]
\nonumber\\
\int{\rm d}\alpha\;s\cdot\nu &=& 2\pi S\cos\theta
[  y/x - (1+y-q^2-x)/2 ]
\label{eq:A8}
\end{eqnarray}
As a final step the scalar products in ${\cal M}_{0,3}^\pm$
are expressed in terms of the variables
$x$, $y$ and $\cos\theta$
\begin{equation}
\begin{tabular}{ll}
$Q\cdot s\; =\; 0$                  & $\qquad\qquad
\ell\cdot s\; =\; -xS\cos\theta/2 $   \\
$Q\cdot q\; =\; (1 + q^2 - y)/2$    & $\qquad\qquad
\ell\cdot q\; =\; (x-y)/2$         \\
$Q\cdot\nu\; =\; (1 - q^2 -x +y)/2$ & $\qquad\qquad
\ell\cdot\nu\; =\; y/2$            \\
$Q\cdot\ell\; =\; x/2$              & $\qquad\qquad
\nu\cdot q\; =\; (1 - q^2 -x)/2$   \\
\end{tabular}
\label{eq:A9}
\end{equation}
and eqs.(\ref{eq:5})-(\ref{eq:8}) follow  immediately.\\
The same steps are performed for the three-body radiative corrections
which are given by the  formula (\ref{eq:A3}), where
\begin{eqnarray}
{\cal M}_{1,3}^+ &=& -\,
\left[\,(1-y/\bar y)^2 + \gamma^2\,\right]^{-1}
\;[\;  q\cdot\nu\; R\cdot\ell\;H_0
\ +\  \epsilon^2\,  Q\cdot\nu\; R\cdot\ell\;H_+
\nonumber\\
&&\ +\  q\cdot\nu\; ( q\cdot\ell\; +\; R\cdot\ell\; Q\cdot q
\;-\; Q\cdot\ell\; R\cdot q )\;H_-
\nonumber\\
&&\ +\ {\textstyle{1\over 2}}\epsilon^2\,
 ( \nu\cdot\ell\; +\; Q\cdot\nu\; R\cdot\ell\;
 -\; R\cdot\nu\; Q\cdot\ell )\;
(H_+ + H_-) \ ]
\label{eq:A10}
\end{eqnarray}
\begin{eqnarray}
H_0 &=&
 4 \left( 1 - p_0 Y_p/p_3  \right)\ln\lambda_G
 + (2p_0/ p_3) \left[ \
{\rm Li}_2\left(  1 - {\textstyle {p_-w_-\over p_+w_+}} \right)
\right.
\nonumber\\
&&\left.
- {\rm Li}_2\left( {\textstyle 1 - {w_-\over w_+}} \right)
- Y_p(Y_p+1) + 2(\ln\epsilon+Y_p)(Y_w+Y_p)\
\right]
\nonumber\\
&& + [ 2p_3 Y_p + (1-\epsilon^2-2y)\ln\epsilon ] /y  + 4
\label{eq:A11}\\
\nonumber\\
H_\pm &=& {\textstyle{1\over 2}}
\left[ 1\pm (1-\epsilon^2)/ y \right] Y_p/p_3
\pm {\textstyle{1\over y}}\ln\epsilon
\label{eq:A12}
\end{eqnarray}

\subsection{Four-body contributions}
Neglecting those terms whose contributions to the decay rates
are ${\cal O}(\lambda_G\ln\lambda_G)$ or smaller
(i.e. vanishing in the limit
$\lambda_G\to 0$) one can cast the contribution of real radiation
into the following expression
\begin{eqnarray}
{\cal M}_{1,4}^+ &=&
{1\over (1-y/\bar y)^2 + \gamma^2 }\,
\left[ {{\cal B}_1^+\over(Q\cdot G)^2}
  - {{\cal B}_2^+\over Q\cdot G\; P\cdot G}
  + {{\cal B}_3^+\over (P\cdot G)^2} \right]
\label{eq:A13}
\end{eqnarray}
where
\begin{eqnarray}
{\cal B}_1^+ &=& q\cdot\nu\;[\ R\cdot\ell\;( Q\cdot G - 1 )
+ \; G\cdot\ell\; R\cdot Q\; - \; Q\cdot\ell\; R\cdot G
\; +\; G\cdot\ell\; R\cdot G \ ]
\nonumber\\
{\cal B}_2^+ &=&
q\cdot\nu\;[\ G\cdot\ell\; R\cdot q\; -\;
q\cdot\ell\; R\cdot G\; +\; R\cdot\ell\;(q\cdot G\; - Q\cdot G\; -
2\, q\cdot Q\;)\ ]
\nonumber\\
&& +\; R\cdot\ell\,(Q\cdot\nu\; q\cdot G\; -\; G\cdot\nu\; q\cdot Q)
\nonumber\\
{\cal B}_3^+ &=&
R\cdot\ell\,( G\cdot\nu\; q\cdot G\; -\; q^2\; P\cdot\nu )
\label{eq:A14}
\end{eqnarray}
The four-body phase space is decomposed as follows
\begin{equation}
{\rm d}{\cal R}_4(Q;q,G,\ell,\nu)  =
{\rm d}z {\rm d}{\cal R}_3(Q;P,\ell,\nu) {\rm d}{\cal R}_2(P;q,G)
\label{eq:A15}
\end{equation}
After $q$ is substituted by $P-G$ integration of
${\cal M}_{1,4}^\pm$
over ${\rm d}{\cal R}_2(P;q,G)$ is performed.
Due to Lorentz invariance
all the integrals which appear in this calculation
can be reduced to the scalar integrals
\begin{equation}
I_n = \int {\rm d}{\cal R}_2(P;q,G) (Q\cdot G)^n
\label{eq:A16}
\end{equation}
Explicit formulae are listed in Appendix C of
\cite{czajk91} for the reduction of tensor integrals
$\int {\rm d}{\cal R}_2(P;q,G) (Q\cdot G)^n G^\alpha$,
$\int {\rm d}{\cal R}_2(P;q,G) (Q\cdot G)^n G^\alpha G^\beta$
and
$\int{\rm d}{\cal R}_2(P;q,G)(Q\cdot G)^n G^\alpha G^\beta G^\gamma$
to the scalar ones.\\
In the next step Dalitz parametrization is employed for
the three-body phase space
${\rm d}{\cal R}_3(Q;P,\ell,\nu)$
in the same way as described in the previous subsection, but
now for $q$ replaced with $P$ in eqs.(\ref{eq:A7})-(\ref{eq:A9}).
In particular $q^2 = \epsilon^2$ is replaced with $P^2 = z$.\\
After integrations over Euler angles $\alpha$ and $\beta$
the contribution of real radiation is splitted into two pieces:
the infrared divergent part of the form
$${\rm Const}\;
[ {\rm F}^\pm_0(x,y) + S\cos\theta\,{\rm J}^\pm_0(x,y) ]\;
I_{div}$$
where
\begin{equation}
 I_{div} = I_{-2} -
(1-y+\epsilon^2)\,I_{-1}/(P\cdot G)
+ \epsilon^2\, I_{0}/(P\cdot G)^2
\label{eq:A17}
\end{equation}
and the rest which is infrared finite.\\
For $\lambda_G\ll 1$  the infrared divergent part
can be integrated over $z$ and the result is proportional to

\begin{eqnarray}
\lefteqn{{1\over\pi}
\int^{z_m}_{\left(\epsilon+\lambda_G\right)}
{\rm d}z\,I_{div} =
  4\left(1 -{p_0\over p_3} Y_p\right)
 \ln\left({z_m-\epsilon^2}\over\epsilon\, \lambda_G \right)
 - 2\ln\left(z_m/\epsilon^2\right) - 2 }
\nonumber\\
&& + {2p_0\over p_3}
\left[
 {\rm Li}_2\left(1-{1-x\over p_+}\right)
+ {\rm Li}_2\left(1-{1-y/x\over p_+}\right)
- {\rm Li}_2\left(1-{1-x\over p_-}\right)
\right.
\nonumber\\
&&\left. \qquad\qquad
- {\rm Li}_2\left(1-{1-y/x\over p_-}\right)
- {\rm Li}_2\left(1-{p_-\over p_+}\right)
- Y_p (Y_p + 1) \right]
\label{eq:A18}
\end{eqnarray}
When this result is added to the contribution of virtual corrections
the infrared divergent terms $\sim\ln\lambda_G$ cancel out.
The sum is simplified using the following identity
\begin{eqnarray}
{\rm Li}_2\left(  1 - {\textstyle {p_-w_-\over p_+w_+}} \right)
- {\rm Li}_2\left( {\textstyle 1 - {w_-\over w_+}} \right)
- {\rm Li}_2\left( {\textstyle 1-{p_-\over p_+} } \right) =
\nonumber\\
{\rm Li}_2\left(w_-\right) - {\rm Li}_2\left(w_+\right)
- 2Y_w\ln p_+
\label{eq:A19}
\end{eqnarray}

\noindent
For the infrared finite part the limit $\lambda_G\to 0$ is performed.
The formulae for the scalar integrals $I_n$ simplify considerably
in this limit and read
$\begin{tabular}{ll}
$ I_{-2} = \pi / (P\cdot G)$  ,
& $\qquad\qquad\qquad
I_{-1} = \pi {\cal Y}_p(z) / \sqrt{-\Delta}$  ,  \\
\end{tabular}$
\begin{equation}
I_n = {\pi(PG)^{n+1}
[\ (Q\cdot P+\sqrt{-\Delta})^{n+1}-(Q\cdot P-\sqrt{-\Delta})^{n+1}\ ]
\over 2(n+1)(P^2)^{n+1} \sqrt{-\Delta}} \quad(n\ge 0)
\label{eq:A20}
\end{equation}
where\footnote{In the present calculation only $n\le1$ is needed.
Thus, instead of eq.(\ref{eq:A20}) one can use:
$ I_n = \pi (P\cdot G)^{n+1}(Q\cdot P)^n/(P^2)^{n+1}\quad$ for
$\quad(n=0,1)$ .}
\begin{eqnarray}
\Delta &=& Q^2P^2 -(Q\cdot P)^2 = -\lambda(z)/4
\nonumber\\
\lambda(z) &=& z^2 - 2(1+y)z + (1-y)^2
\nonumber\\
{\cal Y}_p(z) &=& {\textstyle{1\over 2} \ln\left(
{1-y+z +\sqrt{\lambda(z)}\over 1-y+z -\sqrt{\lambda(z)}} \right) }
\nonumber\\
P\cdot G &=& (z - \epsilon^2)/2
\label{eq:A21}
\end{eqnarray}
Integration of the infrared finite part over $z$
is tedious. It would be
difficult to accomplish this task without FORM~\cite{Form}.
All the integrals which appear can be reduced to
integrals $\int {\rm d}z\,z^n/\lambda^m(z)$  and
$\int {\rm d}z\,z^n\,{\cal Y}_p(z)/\lambda^{m+1/2}(z)$. Thus, they
can be derived from the recursion relations given
in Appendix B of \cite{jk1}.

\subsection{Formulae for ${\rm F}^\pm_1(x,y)$ and ${\rm J}^\pm_1(x,y)$}
After some algebra mentioned in preceding subsections
one derives the following formulae for
${\rm F}^\pm_1(x,y)$:
\begin{equation}
{\rm F}_1^\pm(x,y) =
{\cal H}_1^\pm(x,y)+{\cal H}_2^\pm(x,y,\epsilon^2)
-{\cal H}_2^\pm(x,y,z_m)
\end{equation}
where the first term in r.h.s. is the sum of the three-body
virtual corrections and the infrared divergent four-body ones.
The two other terms originate from the infrared finite four-body
piece. For ${\rm J}^\pm_1(x,y)$ one has:
\begin{equation}
{\rm J}_1^\pm(x,y) =
{\cal K}_1^\pm(x,y)+{\cal K}_2^\pm(x,y,\epsilon^2)
-{\cal K}_2^\pm(x,y,z_m)
\end{equation}
The results of \cite{jk1} and \cite{czajk91} are presented in
this way; see eqs.(3.2)-(3.6) and (4.5)-(4.8) in \cite{jk1},
and (3.1),(A.1)-(A.15) in \cite{czajk91}.\\
Both
${\cal H}_2^\pm(x,y,z)$
and
${\cal K}_2^\pm(x,y,z)$
are given by lenghty expressions:
\begin{eqnarray}
{\cal H}_2^\pm(x,y,z)=
(a^\pm_{1}+za^\pm_{2})
\, {\cal Y}_p(z)/{\lambda}^{3/2}(z)
\,+\, (a^\pm_{3}+za^\pm_{4})\,{\cal Y}_p(z)/\sqrt{\lambda(z)}
\nonumber\\
+{\,}a^\pm_{5}{\,}{\cal Y}_p(z)\sqrt{\lambda(z)}
+{\,}(a^\pm_{6}+za^\pm_{7})/\lambda(z)
\nonumber\\
\,+{\,}a^\pm_{8}{\,}z+a^\pm_{9}{\,}z^{-1}+a^\pm_{10}{\,}\ln{z}
\,+{\,}a^\pm_{11}{\,}
\left[{\rm Li}_2({\cal W}_+(z))+{\rm Li}_2({\cal W}_-(z))\right]
\end{eqnarray}
\begin{eqnarray}
{\cal K}_2^\pm(x,y,z)=
(b^\pm_{1}+zb^\pm_{2})\,{\cal Y}_p(z)/{\lambda}^{5/2}(z)
+{\,}(b^\pm_{3}+zb^\pm_{4})\,{\cal Y}_p(z)/{\lambda}^{3/2}(z)
\nonumber\\
\,+{\,}(b^\pm_{5}+zb^\pm_{6})\,{\cal Y}_p(z)/\sqrt{\lambda(z)}
+{\,}b^\pm_{7}{\,}{\cal Y}_p(z)\sqrt{\lambda(z)}
\nonumber\\
+(b^\pm_{8}+zb^\pm_{9})/\lambda^{2}(z)
+(b^\pm_{10}+zb^\pm_{11})/\lambda(z)
\nonumber\\
\,+{\,}b^\pm_{12}{\,}z+b^\pm_{13}{\,}z^{-1}+b^\pm_{14}{\,}\ln{z}
\,+{\,}b^\pm_{15}{\,}
\left[{\rm Li}_2({\cal W}_+(z))+{\rm Li}_2({\cal W}_-(z))\right]
\end{eqnarray}
where $a^\pm_j$ and $a^\pm_j$ are rational functions of $x$,
$y$ and $\epsilon^2$, and
\begin{equation}
{\cal W}_\pm(z) = \left[\ 1+y-z \pm \sqrt{\lambda(z)}\ \right]/2
\end{equation}
These complicated expressions simplify dramatically for
$z=\epsilon^2$ and $z=z_m$. At these special points
\begin{eqnarray}
&&{\cal Y}_p(\epsilon^2) = Y_p
\nonumber\\
&&\sqrt{\lambda(\epsilon^2)} = 2p_0
\nonumber\\
&&{\cal W}_\pm(\epsilon^2) = w_\pm
\nonumber\\
&&\lambda(z_m) = (x-y/x)^2
\nonumber\\
&&{\cal Y}_p(z_m)/\sqrt{\lambda(z_m)} =
\textstyle{{1\over 2(x-y/x)}\ln\left({1-y/x\over 1-x}\right) }
\nonumber\\
&&{\rm Li}_2({\cal W}_+(z_m))
+ {\rm Li}_2({\cal W}_-(z_m))
= {\rm Li}_2(x) + {\rm Li}_2(y/x)
\end{eqnarray}
and after some algebra eqs.(\ref{F1pm}) and (\ref{J1pm})
are derived.

\vskip0.5cm
{\large\bf Acknowledgements}\\
The authors gratefully acknowledge the fruitful collaboration
with Hans K\"uhn on the subject presented in this article.
M.J. thanks Kacper Zalewski for useful comments.

\end{document}